\documentclass[12pt]{article}

\usepackage{amsmath}
\usepackage{graphicx}
\usepackage{amsfonts}
\usepackage{amssymb}
\usepackage{epsfig}
\usepackage{color}
\usepackage{psfrag}

\newcommand{\EQ}{\begin{equation}}
\newcommand{\EN}{\end{equation}}
\newcommand{\be}{\begin{equation}}
\newcommand{\ee}{\end{equation}}
\newcommand{\bea}{\begin{eqnarray}}
\newcommand{\eea}{\end{eqnarray}}

\newcommand{\re}{{\rm e}}
\newcommand{\rd}{{\rm d}}
\newcommand{\p}{\partial}

\begin{document}
\setcounter{page}{0} \topmargin 0pt
\renewcommand{\thefootnote}{\arabic{footnote}}
\newpage
\setcounter{page}{0}

\begin{titlepage}

\begin{flushright}
SISSA/10/2005/FM
\end{flushright}
\vspace{0.5cm}

\begin{center}
{\large {\bf Mass Generation in Perturbed Massless Integrable Models}}\\
\vspace{1.8cm}
{\large D. Controzzi   
and G. Mussardo}\\
\vspace{0.5cm}
{\em  International School for Advanced Studies and INFN \\
Via Beirut 1, 34100 Trieste, Italy} \\
\end{center}
\vspace{1.2cm}

\renewcommand{\thefootnote}{\arabic{footnote}}
\setcounter{footnote}{0}

\begin{abstract}
\noindent
We extend form-factor perturbation theory to non--integrable deformations 
of massless integrable models, in order to address the problem of mass 
generation in such systems. With respect to the standard renormalisation 
group analysis this approach is more suitable for studying the particle 
content of the perturbed theory. Analogously to the massive case, interesting 
information can be obtained already at first order, such as the identification 
of the operators which create a mass gap and those which induce the confinement 
of the massless particles in the perturbed theory. 
\end{abstract}

\vspace{.3cm}

\end{titlepage}

\newpage

{\bf Introduction.}
Given the large number of remarkable results obtained from the study 
of two--dimensional integrable quantum field theories (IQFTs), at
present one of the most interesting challenges consists of 
developing a systematic approach to study non-integrable models, 
at least when they are deformations of integrable ones. For massive 
field theories a convenient perturbative scheme, based on the exact 
knowledge of the form-factors (FFs) of the original integrable theory,  
was suggested in \cite{dms}. Already at first order, it proved able to provide 
a great deal of information, such as the evolution of the particle content, 
the variation of the masses and the change of the ground state energy 
-- results successfully checked by numerical studies. 

The main purpose of this paper is to extend Form Factor Perturbation Theory (FFPT)
to non--integrable deformations of massless IQFTs. The most fundamental question 
that one may ask in this context is whether a perturbation creates a gap in the 
excitation spectrum --  a problem usually addressed via the renormalisation group 
(RG) equations near a fixed point \cite{RG.book}. Moreover, if massive particles are 
created, one would like to understand whether they are adiabatically related to the 
original massless excitations or, like in the massive case, confinement takes place. 
Since the RG eq.s cannot provide a complete answer to any of the above questions, 
it is worth exploring other alternative routes. The FFPT relies directly on the particle 
description of the unperturbed theory and, for this reason, it seems to be the most natural 
and suitable approach for studying the evolution of the particle content when 
the perturbation is switched on. 

Our analysis is presently limited to the first order of FFPT, its extension to higher orders 
being, as in the massive case, an interesting but non-trivial mathematical problem. Despite 
the fact that one must be careful in handling results at such low order, some 
useful conclusions can nevertheless be reached. For instance, it will be possible to 
discriminate between operators which do not spoil the massless nature of the 
theory and those which instead induce  a mass gap in the spectrum. Moreover, 
the confinement of the original massless excitations can be traced back to the 
non--local properties of the perturbing operator with respect to them. These 
results provide the first information on the perturbed theory and may guide 
a further analysis of its properties. It should be stressed that answering the 
above questions in their full generality is, obviously, a fairly complicated problem since it 
concerns the global structure of the RG flows rather than their local properties 
around the fixed points. It is well known, for instance, that adding a relevant 
perturbation to a massless action does not necessarily imply that the resulting 
infrared theory will be massive: indeed the perturbing operator may induce 
a flow into a new critical point, with some of the massive excitations decoupled from 
the new massless ones \cite{ZAMPHI13,zam.massless}. An example even more subtle is 
given by the roaming trajectories discovered by Al. Zamolodchikov \cite{roamingZam} 
(and further analysed in \cite{roam,ADM}), i.e. an infinite cascade of massless flows 
finally ending in a massive phase.

{\bf FFPT for Massive Field Theories.} Consider non--integrable 
theories obtained as a deformation of an integrable action ${\cal A}_{int}$ 
\EQ
{\cal A} \,=\, {\cal A}_{int} + \lambda \int d^2x \,\Psi(x) 
\,\,\,.
\label{action}
\EN 
The exact knowledge of the FFs of the operator $\Psi(x)$ on the asymptotic 
states of the unperturbed theory allows one to set up an expansion of various 
physical quantities of the new theory in powers of $\lambda$, the so called FFPT 
\cite{dms}. We present initially some known results of FFPT for massive theories 
in a way that is more suitable for the extension to the massless case. It 
is useful to recall that in most of the cases of interest, the integrable 
action of a massive theory can be defined in terms of a deformation of a CFT
\cite{Zam},  
$
{\cal A}_{int} \,=\, {\cal A}_{CFT} + g \,\int d^2x \,\Phi(x) 
$,
where $\Phi(x)$ is a relevant scalar field of conformal weights 
$\Delta_\Phi = \overline \Delta_\Phi < 1$.

Let us first assume that the theory has only one massive particle in the spectrum, 
$A(\beta)$, where $\beta$ parameterises the dispersion relation: 
$p_0=m \cosh \beta\,$ and $p_1 = m \sinh\beta$. The integrability of the 
theory allows one to compute its exact factorised scattering amplitudes \cite{Zam}, 
$S(\beta_{12})~~(\beta_{12} = \beta_1-\beta_2)$, and the FFs \cite{Smirnov} 
of the various operators ${\cal O}$ on the set of asymptotic states 
\EQ
F^{\cal O}
(\beta_1,\beta_2,\ldots,\beta_n) 
\,=\,
\langle 0 \mid {\cal O}(0) \mid A(\beta_1) A(\beta_2) 
\ldots A(\beta_n) \rangle  \,\,\,.
\label{exactFF}
\EN 

A convenient way to study the mass correction induced by the 
non--integrable deformation (\ref{action}) is to employ the 
Hamiltonian formalism, in the same spirit of standard quantum 
mechanics perturbation theory. 
The Hamiltonian associated to (\ref{action}) can 
be written as 
\EQ
H \,=\,\frac{1}{2\pi} \int dx^1 \,T_{00} (x^1,0) \,-\lambda 
\int dx^1 \, \Psi (x^1,0)\,\,, 
\label{Hamiltonian}
\EN
where $T_{\mu\nu}(x)$ is the stress--energy tensor  of the
integrable theory, ${\cal A}_{int}$. The operator 
$T_{00}$ can be expressed in terms of the its trace, $\Theta(x)$, 
using the conservation law $\partial^{\mu} T_{\mu \nu} = 0$,  
\EQ
\label{thetat00}
\partial^2_1 \,\Theta(x^1,x^0) \,= \,(\partial^2_1 - \partial^2_0) 
\, T_{00}(x^1,x^0)\,\,\, . 
\EN
In particular, for the two--particle Form Factor we have 
\begin{eqnarray}
\langle 0 \mid T_{00}(x^1,x^0) 
\mid A(\beta_i)
A(\beta_j)\rangle =
-\sinh^2\frac{\beta_i+\beta_j}{2}\, \, 
\langle 0
\mid \Theta(x^1,x^0) \mid A(\beta_i) A(\beta_j)\rangle 
\,\,\,. 
\label{t00Theta2} 
\end{eqnarray}

The essential results of FFPT are easily re-derived within this formalism.
Let us first consider the unperturbed integrable case, $\lambda=0$.
Evaluating the matrix element of both sides of Eq.~(\ref{Hamiltonian}) 
on the asymptotic states $\langle A(\beta_i)\mid$ and $\mid A(\beta_j)\rangle$, 
and using the relation (\ref{t00Theta2}), one obtains the usual normalisation 
condition for the trace of the stress-energy tensor of the massive integrable theory,
\EQ
\langle A(\beta) \mid\Theta(0)\mid A(\beta) \rangle \,=\,
F^{\Theta}(i \pi) \,=\, 
2 \pi m^2  \,\,\,, 
\label{normalizationTheta}
\EN 
an equation which shows the relationship between the FF of this 
operator and the mass scales of the theory.

Repeating the same procedure for the non--integrable theory (\ref{Hamiltonian}), 
one obtains instead the first order correction to the mass of the particles, as given 
in \cite{dms},
\EQ
\delta m^2 \,\simeq\,2  \,\lambda 
\,F^{\Psi}(i \pi) 
\,\,\,. 
\label{deltam^2}
\EN
If the operator $\Psi(x)$ is non--local with respect to the particles 
$A(\beta)$, $F^{\Psi}(\beta)$ has a pole for $\beta=i \pi$ and 
(\ref{deltam^2}) diverges. This divergent correction to their masses implies 
the confinement of the particles $A(\beta)$, that are no longer excitations 
of the action (\ref{action}) \cite{dms}. This phenomenon appears for instance 
in the magnetic deformation of the low--temperature phase of the Ising model 
\cite{dms,wumccoy,zam-fonseca} as well as in two-frequency sine-Gordon model \cite{dm}.

It should be noticed that, since this is a strong coupling analysis, i.e. 
carried out in the infrared region (IR), if $\lambda$ in (\ref{action}) 
scales under RG, it has to be replaced in (\ref{deltam^2}) by its renormalised value 
at energy of the order of the mass of the theory 
\be
\label{leff}
\lambda \to \lambda^{eff} \simeq \lambda(m^{-1}) \,\,\,.  
\ee 
As a consequence, unless the RG flow is known exactly, quantitative
predictions can be made only on universal mass ratios.

If the theory has $n$ non-degenerate particles, $A_a(\beta)$, with masses $m_a$ 
($a=1,\ldots, n$; $m_a\neq m_{a'}$) the above analysis can be easily extended 
and gives the following mass variation 
\be
\delta m_{a}^2\,\simeq\,  2\,\lambda\,  
\,F_{\bar a a}^{\Psi}(i \pi) 
\,\,\,.
\label{deltam^2.ab}
\ee
When some of the particles, say $n'$ ($n'<n$), have the same mass, this 
equation has to be generalised like in quantum mechanics perturbation
theory for degenerate levels, i.e. the perturbed masses are obtained by 
diagonalising the matrix $\{M_{k,l}\}=\{F^\Psi_{k,l}(i\pi)\}$, where 
indices $k,l$ belong to the degenerate multiplet. If the symmetry of 
the perturbing operator is less than the symmetry of the multiplet, 
the perturbation will typically split it.

{\bf Massless IQFTs.} Massless non--scale invariant IQFTs are associated to RG 
flows between two different fixed points. With respect to their ultraviolet fixed 
point, such theories admit a well--defined description in terms of the corresponding 
CFT perturbed by a relevant operator. However, from the physical point of 
view of selecting the low--energy massless excitations, it is more 
appropriate to view them as irrelevant perturbation of their IR fixed point 
action
\be
{\cal A}_{massless} \,=\, {\cal A}^{IR}_{CFT} + g \,\int d^2x \,
\hat\Phi(x) + \cdots  
\,\,\,, 
\label{actionmassless}
\ee
where the irrelevant field $\hat\Phi(x)$ specifies the approaching direction 
to the CFT of the IR fixed point. The scattering theory of massless IQFTs is 
discussed in detail in \cite{zam.massless} whereas their Form Factors
in \cite{dms2}, so we shall outline only some basic facts below. The
excitations of these theories consist of right ($R$) and left ($L$) moving 
particles. They are defined as $p_1 \geq 0$ and $p_1\leq 0$ branches of the relativistic 
dispersion relation $p_0=|p_1|$, which can be parameterised as $p_0 =
p_1 = (M/2)\, \re^\beta$ for the $R$ movers, $A_R(\beta)$, and $p_0 
= - p_1 = (M/2)\, \re^{-\beta}$ for the $L$ movers, $A_L(\beta)$, where 
$M$ is a mass scale. Within this parameterisation, the Maldenstam variable 
for the $RL$ scattering process is given by: $s_{RL}(\beta_{ij})\, 
= \,M^2 \,\re^{\beta_{ij}}$. Contrary to the massive case, where the 
threshold of the scattering process is given by $\beta_{ij} = 0$, for the $RL$ 
sector of the massless scattering the threshold is reached in the limit $\beta_{ij} 
\rightarrow - \infty$. In the $RR$ and $LL$ sectors the Mandelstam variable is 
always zero, showing that all analyticity arguments of the $S$--matrix theory 
cannot be applied: the scattering amplitudes in these channels can be 
properly defined only as analytic continuation of the massive case
\cite{zam.massless}. For this purpose, in fact, it is useful to regard 
the massless excitations as a particular limit of the massive 
particles\footnote{For instance, if $A(\beta)$ is a massive
excitation of mass $m$ with a $S$--matrix equal to $S(\beta)$, the
massless limit is constructed by shifting the rapidities $\beta \to \beta_{R,L} 
\pm \beta_0/2$ and taking the double limits $\beta_0\to \infty$ and $m\to 
0$ while 
$M = m \re^{\beta_0}$ is kept fixed: $A_{R,L}(\beta) = {\rm lim}_{\beta_0 \to \infty}   
A(\beta\pm\beta_0/2)$. When one considers the S-matrix in the $RR$ and $LL$ 
sectors, the rapidity shifts cancel and therefore $S_{RR}(\beta) = S_{LL}(\beta) = S(\beta)$. 
As functions of the rapidity variable, these amplitudes are then expected to satisfy 
the same equations valid for the massive case.}.  

Writing the $S$--matrix in a compact form  
$
A_{\alpha_1}(\beta_1) A_{\alpha_2}(\beta_2)=  
S_{\alpha_1,\alpha_2}(\beta_{12}) \,
A_{\alpha_2}(\beta_2) A_{\alpha_1}(\beta_1),
\label{masslessSmatrix}
$ $(\alpha_i=R,L$)
the equations satisfied by the FFs can be written in analogy 
to the massive case \cite{dms2}. For the two--particle matrix element 
$F^{\cal O}_{\alpha_1,\alpha_2}(\beta_{12}) = \langle 0|{\cal O}(0)|
A_{\alpha_1}(\beta_1), A_{\alpha_2}(\beta_2) \rangle$, we have for instance
\bea
&&F^{\cal O}_{\alpha_1,\alpha_2}(\beta) \,= \, 
S_{\alpha_1,\alpha_2}(\beta)\, 
F^{\cal O}_{\alpha_2,\alpha_1}(-\beta) \,\,\,;\nonumber \\
&&F^{\cal O}_{\alpha_1,\alpha_2}(\beta+2 \pi i) \,= \, 
\re^{-2 i \pi \gamma_{\cal O}} \,  
F^{\cal O}_{\alpha_2,\alpha_1}(-\beta) \,\,\,,
\eea
where $\gamma_{\cal O}$ is the non-locality index of the operator ${\cal 
O}$ with respect to the massless particles. Their analytic structure, 
however, differs from the massive case. In massive theories, the multi-particle FFs 
are meromorphic functions in the strip $0 \leq {\rm Im} \,\beta < 2\pi$ and present 
simple pole singularities associated either to bound states or to particle-antiparticle 
annihilation processes. In massless theories the same kinds of singularities are expected 
in the $RR$ and $LL$ sectors, since they formally behave like the massive cases. In the 
$RL$ and $LR$ sectors instead, bound state poles are absent while kinematic poles may 
appear only if both particles have vanishing momentum. However, instead of producing a 
recursive equation like in the massive case, here the presence of kinematic poles imposes 
a condition on the asymptotic behaviour of the FF's. In particular, like in the massive 
case, a pole is present in the two-particle FF only if the operator is non-local 
\be
\lim_{\beta_{RL} \to -\infty} F_{RL}^{\cal O}(i\pi +\beta_{RL})=\infty
~~~{\rm for }~{\cal O}~ {\rm non-local}
\label{new-axiom}
\ee
(an analogous equation can be written for the FF in the  $LR$ sector.) 

{\bf FFPT for Massless Field Theories.} Suppose now that an operator $\hat \Psi(x)$ 
of the infrared CFT is added to the effective action (\ref{actionmassless}), so that
its integrability is broken
\be
{\cal A}=\, {\cal A}_{massless} + \lambda \int d^2x \,\hat\Psi(x) 
\,\,\,.
\label{actmaspsi}
\ee 
Repeating initially the analysis of the previous section for the unperturbed 
case $\lambda=0$, one finds\footnote{It should be kept in mind that to avoid trivial 
vanishing of the $RR$ ($LL$) FFs in taking the massless limit of (\ref{t00Theta2}), 
one has to rescale scalar operators by their mass dimension $ {\cal O}(x) \to 
{\cal O}(x)/m^{2\Delta_{\cal O}}$ and define their FFs $F_{\alpha,\alpha}^{{\cal O}}
(\beta_{12}) \,=\, \lim_{m\to 0} \frac{F^{{\cal O}}(\beta_{12})} {m^{2\Delta_{\cal O}}}$, 
where $F^{{\cal O}}(\beta_{12})$ is the two--particle form-factor of the massive 
version of the theory.} the following normalisation conditions for the trace of the
stress-energy tensor
\bea
&&F_{RR}^{\Theta}(i\pi)\,=\,
F_{LL}^{\Theta}(i\pi)=2\pi\,\,\, ;
\label{ThetaRRLL} \\
&&F_{RL}^{\Theta}(\beta) \,=\, 0\,\,\,.  
\label{ThetaRL}
\eea
The last equation can be viewed as an essential property of a massless integrable theory, 
i.e. a non-trivial generalisation to massless non-scale invariant theories of properties 
of CFTs.  

Consider now the case when $\lambda$ is non-zero. If the perturbing operator $\hat\Psi(x)$ 
has vanishing FF on the RL (LR) sector, it is easy to see that, at the lowest order, it does 
not change the masslessness nature of the theory. Indeed, it does not spoil both the validity of  
eq.\,(\ref{ThetaRL}) and the analytic structure of the Green's functions of the original massless 
theory. On the contrary, if the operator $\hat\Psi(x)$ has non-vanishing FFs in the RL (LR) sector 
of the theory, this perturbation immediately generates a mass gap, a quantity which can be estimated 
by sandwiching Eq.~(\ref{Hamiltonian}) on $R$ and $L$ asymptotic states 
\be
\label{deltammassless}
\delta m
\simeq  2 \; \lambda^{eff} \lim_{\beta_{RL}\to- \infty} 
F_{RL}^{\hat \Psi}(i\pi+\beta_{RL}) \,\,\,. 
\ee
From a kinematical point of view, the above limit is the expectation value 
of the perturbing operator at the (zero--energy) threshold of the crossed RL channel. 
In the above equation the effective coupling constant, $\lambda^{eff}$, is defined 
like in (\ref{leff}) with $m\to 0$. As a consequence, if the perturbing operator 
is irrelevant with respect to the IR CFT, $\lambda^{eff}$ scales to zero, i.e. 
the actual mass gap vanishes in the infrared region although it is present 
at intermediate scales. On the other hand, if it is relevant, it will grow to 
the scale of the mass being generated. At this level the relevance of an 
operator has to be established by scaling arguments. In summary, two
conditions have to be fulfilled for generating, at lowest order, a mass gap
in the theory: the perturbing operator has to be relevant with non--vanishing 
RL (LR) matrix elements. 

Moreover, like in the massive case, the mass correction $\delta m$ may be a 
finite or a divergent quantity, depending on the locality properties of the 
perturbing operator with respect to the fields that generate the massless 
particles. If the operator $\hat\Psi(x)$ is local, $\delta m$ is
finite and the massive excitations of the perturbed theory are adiabatically 
related to the massless particles of the original one. If, instead, 
$\hat\Psi(x)$ is a non-local operator, it follows from Eq.~(\ref{new-axiom}) 
and (\ref{deltammassless}) that $\delta m$ diverges: in this case, the original 
massless excitations are confined as soon as $\lambda$ is switched on. 
In other words, the massive particles of the perturbed theory are not 
associated, in this case, to the operators that create the original massless 
ones. The examples discussed below should help in clarifying these two situations. 

If the theory contains more than one type of massless particle $A_{a,R/L}$ 
($a=1,\ldots,n$) the previous approach has to be generalised in analogy 
with perturbation theory for degenerate levels. Since the particles
$A_{a,\alpha}$ form a complete basis for the scattering theory, by using FFPT it 
should be possible, in principle, to predict whether {\em any} massless excitations 
survive, and this is a clear advantage with respect to the RG. Although 
a complete answer to this question involves the entire series in $\lambda$, 
nevertheless the first order of FFPT may provide useful hints on the decoupling of 
massive and massless modes of the theory under investigation.

{\bf Massless flows between minimal models.} Let us now apply the
above methods to some specific examples, starting from the massless
flow between the Tricritical Ising Model (TIM) and the Critical Ising 
Model (CIM). The quantum field theory associated to this RG flow can 
either be seen as TIM perturbed by its sub-leading energy operator $\epsilon'$ 
of conformal dimensions $\Delta_{\epsilon'} = \bar \Delta_{\epsilon'}=3/5 $ 
or as CIM perturbed by the irrelevant operator $T\bar T$ (see \cite{zam.massless} 
and references therein). The factorised scattering theory for this massless 
flow was first proposed in \cite{zam.massless} and the basic FFs calculated 
in \cite{dms2}. The spectrum consists of massless neutral fermions, 
with S-matrix $S_{RR}(\beta)=S_{LL}(\beta)=-1$, while $S_{RL}(\beta) 
= \tanh \left(\beta/2-i\pi/4 \right )$. As well as studying 
the non-integrable theory obtained by the insertion of the energy operator 
$\epsilon(x)$ of the CIM, we will also consider the deformation of the 
massless action by the disorder operator $\mu(x)$. The latter is 
non-local, $\gamma_{\mu} = 1/2$, with respect to the massless fermion 
excitations. The energy operator has two particle FFs only in the $RL$ 
sector of the form \cite{dms2}
\be
F^\epsilon_{RL}(\beta)= Z_\epsilon 
\exp \left ( \frac{\beta}{4}-\int \frac{\rd t}{t}
\frac{\sin^2 \left ( \frac{(i\pi - \beta)}{2 \pi} \right )}
{\sinh t \cosh \frac{t}{2} } \right ) \,\,\,,
\label{epsilon}
\ee
where $Z_\epsilon$ is a normalisation constant. The disorder operator, 
on the other side, has also FFs in the $RR$ and $LL$ sectors, of the 
form $F_{RR}^{\mu}(\beta) = Z'_{\mu} \tanh(\beta/2)$. 
For the FFs in the $RL$ sector we find
\bea
F_{RL}^{\mu}(\beta) = Z_{\mu}
\exp \left (- \frac{\beta}{4}-\int \frac{\rd t}{t}
\frac{\sin^2 \left ( \frac{(i\pi - \beta)}{2 \pi} \right )}
{\sinh t \cosh \frac{t}{2} } \right )\,\,\,. 
\label{mu-RL}
\eea
The above results agree with the roaming limit of the FFs of 
the sinh-Gordon theory \cite{sh-Gordon} -- a limit in which the 
sinh-Gordon model corresponds to the above massless flow
\cite{roamingZam,dms2,ADM}. 

Using now (\ref{deltammassless}) and (\ref{epsilon}), it is easy to see 
that the perturbation by $\epsilon(x)$ induces a finite mass in the 
system, as it could have been expected on different grounds. At the 
critical point, in fact, $\epsilon(x)$ is bilinear in the fermionic 
operators that generate the massless particles, $\epsilon \sim \bar{\psi} 
\psi$, and therefore the perturbed theory describes massive fermions in 
the presence of an irrelevant perturbation $T\bar T$. 

Consider now the perturbation of the massless action by the non-local 
operator $\mu(x)$. By computing the limit (\ref{deltammassless}) of 
the two--particle FF (\ref{mu-RL}) of this operator, one sees that 
in this case $\delta m$ diverges, i.e. the initial excitations can 
no longer propagate as asymptotic states in the new vacuum of the 
theory created by the insertion of this field. Like in the 
massive case, there is a simple explanation of this confinement 
phenomenon in terms of the LG effective description of the theory 
\cite{dms}. Indeed, in the unperturbed theory the elementary 
excitations can be equivalently considered as massless kinks interpolating 
between two degenerate minima of the LG potential \cite{fsz}. However the 
insertion of the disorder magnetic operator $\mu(x)$ lifts the degeneracy 
between the minima, thus making the kinks unstable.

As a matter of fact, the flow between the TIM and the CIM is the simplest example 
of a one-parameter family of RG trajectories interpolating between the conformal 
minimal models ${\cal A}_p$, with central charge $c_p=1-6/p(p+1)$ ($p=3, 4$ describe 
the CIM and TIM respectively). The flows start from ${\cal A}_\infty$ and
pass close to all the other minimal models, remaining massless all the way 
down to the very last fixed point, $p=3$, after which they become massive 
\cite{roamingZam,roam,ADM}. The trajectories going out from each critical 
point are described as ${\cal A}_p$ perturbed by the operator $\phi_{13}$ \cite{ZAMPHI13}: 
${\cal A}^{eff}_p = {\cal A}_p +\lambda \int \rd^2 x \phi^p_{13}$ (where the 
upper index in $\phi_{13}$ indicates the relative CFT) and the excitations are 
massless kinks interpolating between the $(p-2)$ degenerate vacua of the 
effective LG potential \cite{fsz}. An interesting problem consists of predicting
the evolution of the spectrum along these flows, in particular the 
successive decoupling of the massless modes in the cascade of massless
RG flows, by applying FFPT. The analysis of this problem is however beyond 
the scope of the present letter. 

{\bf Spinon confinement in sigma models.}
Another important application of FFPT is in the study of the 
mass spectrum of the $O(3)$ non-linear sigma-model with a topological 
term
\be
{\cal A} _\theta =\frac{1}{2f^2} \int \rd^2 x \left 
( \p _\mu n_\alpha \right )^2 + i\, \theta \,T, ~~~(\alpha=1,2,3 \; ;~ n_\alpha^2=1)
\label{o3.1}
\ee
where $f$ and $\theta$ are dimensionless coupling constants and $T$
is the integer-valued topological term related to the instanton solutions 
of the model. The two values $\theta = (0, \pi)$ are the only ones 
for which the action (\ref{o3.1}) is known to be integrable. At 
$\theta = 0$ the excitations form a massive $O(3)$ triplet whose 
scattering theory was constructed in \cite{Zam,wiegmann}. At $\theta = \pi$ 
the theory is instead massless \cite{zam-zam.massless,ah,shankar.read,merons} 
and corresponds to the RG flow between the $c=2$ CFT and the $SU(2)_1$ 
Wess-Zumino-Witten (WZW) model. The factorised scattering theory was 
suggested in \cite{zam-zam.massless}: it consists of right and left 
doublets, $A_{a,R}$ and $A_{a,L}$ ($a=1,2$), that transform according to 
the $s = 1/2$ representation of $SU(2)$ (spinons). However, as soon as 
one moves away from $\theta = \pi$, the spinons confine \cite{affleck,cm} 
and the actual spectrum of the theory in the vicinity of this point has 
been determined in \cite{cm}. Let us discuss in some detail how 
the spinon confinement takes place. In the FFPT this amounts to show 
that the topological term is non--local wrt the fields that create the 
spinons, a property that can be easily checked by looking at the CFT 
limit of these operators. 

Consider ${\cal A}_\pi$ as our unperturbed IQFT. Close to the IR fixed point the massless flow can be 
described as a $SU(2)_1$ WZW model perturbed by the marginally irrelevant perturbation $({\rm Tr} \,g )^2$ 
\cite{zam-zam.massless,ah,shankar.read}, ${\cal A}^{eff}_\pi = {\cal A}_{SU(2)_1} + \gamma \int \rd ^2 x 
({\rm Tr} \,g )^2\,\,\, $ ($ \gamma>0$), where $g$ is the SU(2) matrix field. In terms of this formulation 
the perturbation that moves the topological term away from $\theta = \pi$ is proportional to ${\rm Tr} \,g$ 
\cite{ah}, i.e. to the only relevant $SU(2)$ invariant operator in the theory that breaks parity. Thus, in 
the vicinity of $\theta = \pi$, the model is described by the effective action \be {\cal A}^{eff} ={\cal 
A}_{\pi}^{eff} + \eta \int \rd ^2 x {\rm Tr} \,g \, , \label{o3.3} \ee where $\eta$ is a 
function\footnote{The form of this function determines the 
dependence on $(\theta-\pi)$ of the mass gap $m$, since it scales as $m \sim \eta^{2/3}$, up to 
logarithmic correction \cite{agsz}. In a recent paper \cite{Gorsky} it has been suggested 
that the gap behaves like $(\theta-\pi)^{1/2}$, 
which would imply a dependence of 
$\eta$ on $(\theta-\pi)$ that it is not linear, as usually assumed.} 
of $(\theta - \pi)$ that vanishes when $\theta = \pi$. 

As discussed in Ref.~\cite{spinons}, the spinons are created
by the primary operator $\phi^\pm (z)$ ($\bar\phi^\pm(\bar z)$) 
with scaling dimension $h=(1/4,0)$ $(\bar h=(0,1/4))$. They enter 
the operator product expansion (OPE)   
\bea
\phi^\alpha(z)\phi^\beta(w)=(-)^q (z-w)^{-1/2}\epsilon^{\alpha \beta}
\left (1+1/2 (z-w)^2 T(w) \ldots \right )
\nonumber \\ 
-(-)^q (z-w)^{1/2}(t_a)^{\alpha \beta}
\left (J^a(w)+1/2(z-w)\p J^a(w) \ldots  \right ),
\label{spinon.OPE}
\eea
where $\epsilon^{+-}=-\epsilon^{-+}=1$, $(t_a)^{\alpha \beta}$ are the
generators of the algebra, and $q$ takes the values $q=0$ for states that 
are created by an even number of spinons and $q = 1$ if the number of spinons is 
odd. The OPE between the spinon operator and the $SU(2)$ currents $J^a(z)$ 
is standard: $J^a(z) \phi^\alpha(w) = (t^a)^\alpha_{~\beta} \phi^\beta(w)/(z-w) + 
\ldots ~.$ From these OPEs it follows that $J^a$ and $\phi^\alpha$ are 
mutually local while $\phi^\alpha$ and $\phi^\beta$ are not. In fact 
taking $\phi^\alpha(z)$ around $\phi^\beta(w)$ by sending $z\to z\,
\re^{2\pi i}$ it produces a factor  $\re^{2\pi i \,\gamma_{\phi}}$ with 
$\gamma_{\phi} = 1/2$. 

Therefore $({\rm Tr} \,g)^2 \simeq \bar J^a J^a$ is local with respect to 
the spinons and this explains why they are the fundamental excitations of 
${\cal A}_\pi$, regardless of whether the perturbation is marginally relevant 
or irrelevant. Since ${\rm Tr} \,g $ is proportional to ($\phi^+ \bar \phi^- +\phi^-
\bar \phi^+ $), the OPE (\ref{spinon.OPE}) implies that this operator 
is instead non-local with respect to the spinons. Hence they get 
confined as soon as the operator ${\rm Tr}\, g$ is added to ${\cal 
A}_\pi$, i.e. the perturbed model (\ref{o3.3}) has no longer spin $1/2$ 
excitations. As discussed in \cite{cm}, the actual massive excitations of the 
O(3) sigma model with $\theta$--term consists of a triplet of particles and 
a singlet, the former stable for all value of $\theta$ whereas the latter 
stable only in an interval of values of $\theta$ near $\theta = \pi$.  

\bigskip

We are grateful to G. Delfino for important discussions and A. Nichols
for reading the manuscript. D.C would also like to thank A. Tsvelik and 
K. Schoutens for useful discussions. This work is within the activity of 
the European Commission TMR program HPRN-CT-2002-00325 (EUCLID).


\begin{thebibliography}{99}


\bibitem{dms} G. Delfino, G. Mussardo and P. Simonetti,  Nucl.Phys. {\bf B 
473} 
(1996) 469. 
\bibitem{RG.book}see for instance, 
A.O. Gogolin, A.A. Nersesyan and A.M. Tsvelik, {\it
Bosonization in Strongly Correlated Systems}, Cambridge University Press, 1999.
\bibitem{ZAMPHI13} A.B. Zamolodchikov, Sov. J. Nucl. Phys. {\bf 48}
(1987), 1090.
\bibitem{zam.massless} Al.B. Zamolodchikov, Nucl.Phys. {\bf B 358} (1991) 
524.
\bibitem{roamingZam} Al.B. Zamolodchikov, {\em Resonance Factorized 
Scattering and Roaming Trajectories}, ENS-LPS-335 (1991).
\bibitem{roam} 
M.J. Martins, Phys. Rev. Lett. {\bf 69} (1992), 2461; 
M.J. Martins, Phys. Lett. {\bf B 304} (1993), 111; 
P. Dorey and F. Ravanini, Nucl. Phys. {\bf B 406}  (1993), 708; 
P. Dorey and F. Ravanini, Int. J. Mod. Phys. {\bf A 8} 
(1996), 873. 
\bibitem{ADM} C. Ahn, G. Delfino and G. Mussardo, Phys. Lett. {\bf B 317}, 
573 (1993).
\bibitem{Zam} A.B. Zamolodchikov and Al.B. Zamolodchikov,   
Ann. Phys. NY {\bf 120} (1979) 253; A.B. Zamolodchikov, 
Adv. Stud. Pure Math. {\bf 19} (1989), 641.
\bibitem{Smirnov} F. A. Smirnov, {\it Form Factors in Completely 
Integrable Models of Quantum Field Theory}, World Scientific, 
Singapore (1992); M. Karowski and P. Weisz, Nucl. Phys. {\bf D139}, 455 (1978),
B. Berg, M. Karowski, P. Weisz, Phys. Rev. {\bf D19}, 2477 (1979).
\bibitem{wumccoy}B.M. McCoy and T.T. Wu, Phys. Rev. {\bf D 18} (1978), 1259.
\bibitem{zam-fonseca}P. Fonseca and A.B. Zamolodchikov, J. Stat. Phys,
{\bf 110} (2003), 527. 
\bibitem{dm} G. Delfino and G. Mussardo, Nucl. Phys. {\bf B 516}, 675 
(1998).
\bibitem{dms2} G. Delfino, G. Mussardo and P. Simonetti, Phys. Rev. 
{\bf D 51} (1995) 6622.
\bibitem{fsz}P. Fendley, H. Saleur and Al.B. Zamolodchikov,
  Int. J. Mod. Phys. {\bf A 8} (1993), 5751. 
\bibitem{sh-Gordon}A. Fring, G. Mussardo and P. Simonetti, Nucl. Phys. 
{\bf B 393}(1993) 413; A. Koubek and G. Mussardo, Phys. Lett. 
{\bf B 311} (1993), 193; M. Lashkevich, hep-th/94061118.
\bibitem{wiegmann}P. Wiegmann, Phys. Lett. {\bf B 152}(1985) 209.
\bibitem{zam-zam.massless} A.B. Zamolodchikov and Al.B. Zamolodchikov,   
Nucl.Phys. {\bf B 379} (1992) 602.
\bibitem{ah} I. Affleck and F.D.M. Haldane, Phys. Rev. 
{\bf B 36} (1987) 5291.
\bibitem{shankar.read}
R. Shankar and N. Read, Nucl.Physics. {\bf B 336} (1990)457.
\bibitem{merons}W. Bietenholz, A. Polchinsky and U.-J. Wiese, 
Phys. Rev. Lett. {\bf 75} (1995), 4524.
\bibitem{affleck}I. Affleck, in {\em Dynamical Properties of unconventional
  magnetic systems}, Kluwer Academic Publishers, cond-mat/9705127. 
\bibitem{cm} D. Controzzi and G. Mussardo, Phys. Rev. Lett. {\bf 92}
(2004) 21601.
\bibitem{agsz} 
I. Affleck, D. Gepner, H.J. Schulz and T. Ziman, {\em J. Phys. }
{\bf A 22}(1989), 511.
\bibitem{Gorsky} A. Gorsky,  M. Shifman and  A. Yung, hep-th/0412082.
\bibitem{spinons}P. Bouwknegt, A.W.W. Ludwig and K. Schoutens, Phys. Lett. 
{\bf B 338}, 448 (1994);D. Bernard, V. Pasquier and D. Serban, Nucl. Phys {\bf
B 428} (1994), 428; K. Schoutens, Phys. Rev. Lett {\bf 79}, 2608
(1997).

\end{thebibliography}
\end{document}